\documentclass[aps,pra,epsfigure,twocolumn,showpacs]{revtex4}
\usepackage{dcolumn}    
\usepackage{bm} 
\usepackage{graphicx}
\usepackage{amsmath}    
\usepackage{latexsym}
\usepackage{amsfonts}   
\usepackage{amssymb}
\usepackage{array}      
\usepackage{epsfig}
\usepackage{epstopdf}
\usepackage{times,txfonts}

\newcommand{\ket}[1]{\left\vert#1\right\rangle}

\newcommand{\nbar}{\overline{n}}

\begin{document}
\title{Critical assessment of two-qubit post-Markovian master equations}
\author{S. Campbell$^{1,2,3}$, A. Smirne$^4$, L. Mazzola$^{1,5}$, N. Lo Gullo$^2$, 
B. Vacchini$^4$, Th. Busch$^{2,3}$, 
and M. Paternostro$^1$}
\affiliation{
$^1$Centre for Theoretical Atomic, Molecular, and Optical Physics, School of Mathematics and Physics, Queen's University, Belfast BT7 1NN, United Kingdom\\
$^2$Department of Physics, University College Cork, Republic of Ireland\\ 
$^3$Quantum Systems Unit, Okinawa Institute of Science and Technology, Okinawa, Japan\\
$^4$Dipartimento di Fisica, Universit\'a degli Studi di Milano, Via Celoria 16, I-20133 Milan, Italy \& INFN, Sezione di Milano, Via Celoria 16, I-20133 Milan, Italy\\
$^5$Turku Centre for Quantum Physics, Department of Physics and Astronomy, University of Turku, FI-20014 Turun yliopisto, Finland}

\begin{abstract}
A post-Markovian master equation has been recently proposed as a tool to describe the evolution of a system coupled to a memory-keeping environment [A. Shabani and D. A. Lidar, Phys. Rev. A {\bf 71}, 020101 (R) (2005)]. For a single qubit affected by appropriately chosen environmental conditions, the corresponding dynamics is always legitimate and physical. Here we extend such situation to the case of two qubits, only one of which experiences the environmental effects. We show how, despite the innocence of such an extension, the introduction of the second qubit should be done {\it cum grano salis} to avoid consequences such as the breaking of the positivity of the associated dynamical map. This hints at the necessity of using care when adopting phenomenologically derived models for evolutions occurring outside the Markovian framework.
\end{abstract}
\date{\today}
\pacs{03.65.Yz, 03.67.-a, 42.50.Lc } 
\maketitle

\section{Introduction}
The study of open quantum systems, which has been extensive over the years~\cite{Breuer,Weiss}, is interesting for many stimulating reasons. First, the occurrence of uncontrolled system-environment interactions constitutes a fundamental obstacle for the realization of reliable control over quantum devices, and the challenge is to design practical schemes for quenching such unwanted couplings.
Second, more fundamentally, the dynamics of open quantum systems offer a way to access and explore the way genuine quantum features are smeared out into classical ones~\cite{ZurekRMP03}. 

Very often, the formal analysis of an open quantum dynamics is performed by invoking the use of two simplifications:  the weak system-environment coupling and the forgetful nature of the environmental system~\cite{LindbladCMP76,GoriniJMP76}. This defines the so-called Markovian framework, which is often useful for the grasping of a qualitative understanding of a system-environment evolution and sometimes even physically justified by  particularly favorable working conditions.
Notwithstanding its pragmatic handiness, it should be kept in mind that Markovianity is only an approximation and the conditions for its application frequently do not match the reality of a given physical situation~\cite{YablonovitchPRL91}. The increasing awareness of the limitations in the Markovian framework, the identification of explicit cases of non-Markovian system-environment evolution and the ability to experimentally simulate structured reservoirs with inherent non-trivial dynamics~\cite{BiercukNat09} have triggered the study of open quantum systems beyond the Markovian regime~\cite{BellomoPRL07}.
A considerable number of techniques, both analytic and numerical, have been put forward with the goal of accurately tackle non-Markovian evolution~\cite{Breuer,Nakajima,BreuerPRA99,GarrawayPRA97,PiiloPRL08}.
Among them, post-Markovian (postM) master equations have been proposed as a means to interpolate between fully Markovian Lindblad-like approaches and the exact Kraus-operator picture of the reduced dynamics of a system coupled to its environment~\cite{YuPLA99,ShabaniPRA05}, offering clear conditions for physicality of the dynamics. This circumstance is particularly relevant if one considers that the prerequisites for meaningful evolution (i.e. the dynamical map evolving the density operator has to be trace-preserving, positive and completely positive) are  often violated by dynamics arising from non-Markovian master equation.

In this paper we show how even a well-tested tool as the post-Markovian master equation retains strong limit of applicability hidden in it. 
We consider a qubit system embedded in an environment whose dynamics is meaningfully described by a post-Markovian master equation and an ancillary qubit, completely decoupled from the system and the environment, whose unitary evolution is ruled by a von Neumann equation. As the evolutions of system and ancilla are independent, the master equation of the enlarged system containing the system and the ancilla can be naturally written as the sum of the generators of the two disjointed dynamical evolutions. We show how the solution of such a master equation breaks positivity and creates correlations between the two non-interacting qubits even when initially prepared in a factorized state. Following the formulation of Shabani and Lidar, we go beyond the intuitive approach and derive a post-Markovian master equation which accounts {\it ab initio} of the ancillary degree of freedom. Such a master equation for the extended two-qubit systems fails to describe a completely positive or even only positive dynamics, as in the previous case. The loss of positivity of the two extended evolutions occurs under the same conditions for which physical dynamics of the single qubit-system is guaranteed~\cite{ManiscalcoPRA06}. As such our results shed light on the limit of validity of post-Markovian master equations and warn against the hazard of naive extensions of non-Markovian equations of motion. 

The remainder of the paper is organized as follows. Section II outlines the postM master equation and its derivation that will be focus of this paper. Section III deals with an intuitive approach to introducing an ancillary system into the evolution and examines the loss of physicality in the map. In Section IV we employ the methods of Shibani-Lidar and derive the associated postM master equation explicitly including the ancilla in the derivation and show this also fails to give a physical map. Section V we discuss the steps necessary to resolve the loss of physicality when introducing an ancillary system. Section VI we draw some final conclusions. Finally, there is an Appendix in which some of the more technical steps involved in the calculations are presented. 

\section{Post-Markovian master equation}
It is instrumental for the second part of our analysis and pedagogically quite useful to briefly remind of the key steps needed in order to derive the postM master equation proposed in~\cite{ShabaniPRA05}. The basic observation is that the operator-sum representation of a system-environment dynamics corresponds to making a single measurement on the environment at time $t$. In such a measurement picture, a master equation in the Lindblad form can be seen as a continuous series of measurements performed at infinitely closely spaced times~\cite{ShabaniPRA05}. The postM approach interpolates such extremal cases: At time $t=0$, system and environment start their joint evolution until, at a random time $t'{<}t$, a measurement on the bath is performed and the state of the system becomes $\Lambda(t'){\bm \rho}(0)$, where $\Lambda(t')$ is a one-parameter map independent of the final time $t$. The measurement resets the state of the environment while the joint system-environment evolution proceeds from then until time $t$, when the final measurement is performed.
The probability for the extra environmental measurement  at $t'$ to occur depends on the memory properties of the environment itself, which are accounted for by the introduction of a memory kernel $k(t',t)$. The final state of the system is thus obtained averaging out over the different random times as ${\bm \rho}(t)=\sum_{random\ t'}k(t',t)\Lambda(t-t'){\bm \rho}(t')$. 

The derivation proceeds with the discretization of the interval of time $[0,t]$ into segments of equal length $\epsilon$. By evaluating the difference of the density matrix operator at two consecutive instant of times, dividing for $\epsilon$ and performing the limit for $\epsilon{\rightarrow}0$ we obtain the master equation 
\begin{equation*}
\dot{\bm \rho}{=}\int_0^t dt' \left[ k(t',t) \Lambda(t'){\partial}_{(t-t')}+{\partial_t k(t',t)}\Lambda(t')\right]{\bm \rho}(t-t').
\end{equation*}
In the spirit of perturbation theory, the relation ${\bm \rho}(t-t')=\Lambda(t-t'){\bm \rho}(0)$ is used inside the integral (meaning that in the first order the map describing the dynamics belongs to a one-parameter family). Assuming that the inverse of the map ${\Lambda(t-t')}$ exists, one can write ${\bm \rho}(0)=\Lambda^{-1}(t-t'){\bm \rho}(t-t')$ and, with the assumption that the memory kernel function depends on one parameter only $k(t',t)=k(t')$, the master equation above becomes
\begin{equation}
\label{ShabLidOrig}
\frac{\partial{\bm \rho}(t)}{\partial t}=\int_0^t dt' k(t')\Lambda(t') \frac{\partial \Lambda(t-t')}{\partial (t-t')}\Lambda^{-1}(t-t') {\bm \rho}(t-t').
\end{equation}
In order to obtain a master equation for each particular problem, one needs to specify the form of the map $\Lambda(t)$. In the case of a Markovian super-operator $\Lambda(t')=e^{\mathcal{L}t'}$ the postM master equation attains the form
\begin{equation}
\label{SLeq}
\frac{d{\bm \rho}(t)}{dt}=\hat{\cal L}\int^t_0 dt' k(t')e^{\mathcal{L}t'}{\bm \rho}(t-t').
\end{equation}
We consider here a qubit interacting in a dissipative fashion with a non-zero temperature bosonic bath, the Markovian Liouvillian $\hat{\cal L}$ being in the form
\begin{equation}
\begin{aligned}
\hat{\cal L}{\bm \rho}&{=}\gamma\nbar[\hat{{\bm \sigma}}_+{\bm \rho}\hat{{\bm \sigma}}_-{-}\frac{1}{2}\{\hat{{\bm \sigma}}_-\hat{{\bm \sigma}}_+,{\bm \rho}\}]
+\gamma(\nbar+1)[\hat{{\bm \sigma}}_-{\bm \rho}\hat{{\bm \sigma}}_+{-}\frac{1}{2}\{ \hat{{\bm \sigma}}_+\hat{{\bm \sigma}}_-,{\bm \rho}\}]
\end{aligned}
\end{equation}
where $\hat{{\bm \sigma}}_{\pm}$ are the spin raising and lowering operators. The memory function of the bath $k(t)$ is taken of exponential form $k(t)=\chi\text{exp}(-\chi t)$. 
The map $\hat\Phi(t)$ that arises from  Eqs.~(\ref{SLeq}) with such a choice of the Liouvillian super-operator was considered extensively in Refs.~\cite{ManiscalcoPRA06} and was demonstrated to be completely positive for all range of parameters. Therefore the master equation in Eq.~(\ref{SLeq}) gives a physically meaningful evolution for a single damped qubit. 

\begin{figure*}
 \includegraphics[scale=0.6]{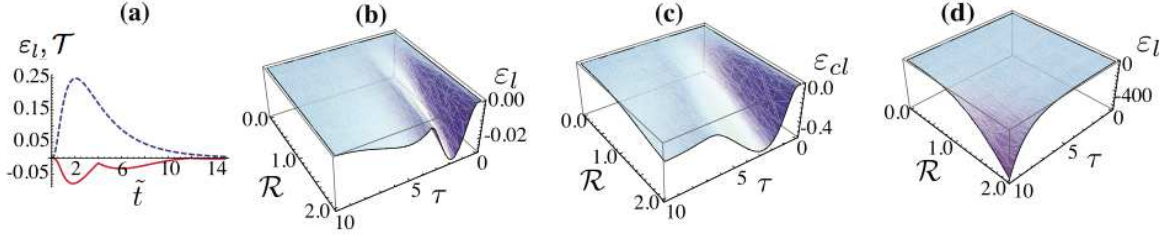} \caption{(Colors online)
 {\bf (a)} Trace distance $\mathcal{T}$ (dashed line) and lowest eigenvalue $\varepsilon_l$ (solid line) of the mapped matrix ${\bm \rho}(t)$ solution of Eq.~(\ref{model}) against the rescaled interaction time $\tilde t={J}t$ for the initial state ${\bm \rho}(0)=\ket{0}_S\otimes(\ket{0}+\ket{1})_A/{\sqrt{2}}$ and $J=2$. {\bf (b)} Lowest eigenvalue $\varepsilon_l$ of the matrix ${\bm \rho}(t)$ as a function of rescaled time $\tau=\gamma t$ and the ratio ${\cal R}=J/\gamma$ for the same initial state of panel {\bf (a)}. {\bf (c)} Plot of the lowest eigenvalue $\varepsilon_{cl}$ of the Choi matrix associated with the dynamics of ${\bm \rho}(t)$. {\bf (d)} Lowest eigenvalue $\varepsilon_l$ of the matrix ${\bm \rho}(t)$ solving Eq.~(\ref{pulitaL}). In all the panels we have taken $\nbar=1$ with $\chi{=}\gamma$.}
 \label{figura}
 \end{figure*}

\section{Introducing an ancillary system: Intuitive approach} 
Here we consider the extended system comprising a damped qubit, whose evolution follows the postM master equation in Eq.~(\ref{SLeq}), and an ancillary qubit evolving in time unitarily and independently from the first qubit.
The introduction of ancillary degrees of freedom is useful in many applications, for example in view of quantifying the degree of non-Markovianity of the dynamics of the system \cite{RivasPRL10} (this can be done by monitoring the evolution of the entanglement between system and ancilla) and for inferring the specifics of the system-environment interaction \cite{CampbellPRA10} (one can access the property of an environment affecting an inaccessible system by means of continuous measurements on an ancilla).

To derive the master equation describing the extended system we use the following line of reasoning. The starting point of a microscopic approach is the von Neumann equation for the density operator of system, ancilla and environment $\dot{{\bm \rho}}_{SAE}=-i\left[\hat H_{SE}\otimes\hat\openone_A+\hat H_A\otimes\hat\openone_{SE},{\bm \rho}_{SAE}\right]$, containing a Hamiltonian term coupling the system and the environment $H_{SE}$, from which decoherence of the system arises, and an independent ancillary Hamiltonian term $\hat H_A$. By tracing out the environmental degrees of freedom one obtains 
$$\dot{{\bm \rho}}_{SA}=-i\mathrm{Tr}_E\left[\hat H_{SE}\otimes\hat\openone_{A},{\bm \rho}_{SAE}\right]-i\left[\hat H_A\otimes\hat\openone_{S},{\bm \rho}_{SA}\right],$$ meaning that the generator of the environment-induced system evolution and the one of the ancillary evolution simply sum up. As the postM generator accounts for the description of the dynamics of the single damped qubit, the most natural way to construct the master equation for the two qubits is to substitute the term corresponding to the dissipative dynamics with the postM generator of Eq.~(\ref{SLeq}), thus obtaining
\begin{equation}
\label{model}
\frac{d{\bm \rho}(t)}{dt}{=}{-}i[\hat H,{\bm \rho}(t)]{+}\hat{\cal L}_S\otimes\hat\openone_A\!\!\int^t_0\!\!dt' k(t')e^{\hat{\cal L}_S t'}\!\!\otimes\hat\openone_A{\bm \rho}(t-t'),
\end{equation} 
with $\hat H=\hat H_A\otimes\hat\openone_S$ and ${\bm \rho}(t)$ the density matrix for the system-ancilla pair. A first check supporting the reasonableness of the previous equation is that when tracing out over the ancilla (system) degree of freedom we obtain the postM equation of the damped system qubit (von Neumann equation for the ancilla). In order to be meaningful this master equation has to be trace preserving and has to give rise to a physical dynamics, meaning that ${\bm \rho}(t)$ has to be a density matrix (trace one, Hermitian and positive) at any time and the map associated has to be completely positive. We choose the following form for the ancillary Hamiltonian
\begin{equation}
\label{HA}
\hat H_A=J \hat{{\bm \sigma}}_z^A,
\end{equation}
where $J$ is the frequency split between the logical states of the ancilla and $\hat{{\bm \sigma}}_z^A$ is the corresponding $z$-Pauli operator.
In order to solve Eq.~(\ref{model}) we decompose ${\bm \rho}(t)$ in terms of the damping basis~\cite{BriegelPRA93} of $\hat{\cal L}$, as 
\begin{equation}
{\bm {\bm \rho}}(t)=\sum_{i,j}c_{ij}(t) \hat{{\bm \sigma}}_i^S\otimes\hat{{\bm \sigma}}_j^A~~~~~~~(i,j=0,z,\pm)
\end{equation}
where $\hat{{\bm \sigma}}_i^{A,S}$ are the right eigen-operators of $\hat{\cal L}_S$ such that $\hat{\cal L}_S\hat{{\bm \sigma}}_i^{S}=\lambda_i\hat{{\bm \sigma}}_i^{S}$ [with $\hat{{\bm \sigma}}_0=(1/2)(\hat\openone-\hat{{\bm \sigma}}_z/(2\nbar+1))$] and $c_{ij}(t)$ are time-dependent coefficients. 
The choice of such an operator basis for the density matrix is very convenient, as the action of $\hat{\cal L}_S$ on each element of the decomposition of ${\bm \rho}(t)$ simply depends on the eigenvalues associated with the damping operators. Moreover, it turns out that the Hamiltonian of Eq.~(\ref{HA}) is diagonal with respect to the damping basis.
By introducing the dual of the damping basis~\cite{BriegelPRA93} (defined as the solution of the orthonormality condition $\text{Tr}[\check{\hat{{\bm \sigma}}}_i \hat{{\bm \sigma}}_j]=\delta_{ij}$, with $\delta_{ij}$ the Kronecker delta function) it is possible to obtain a set of 16 equations for the time-dependent coefficients $c_{ij}(t)$. These equations can then be solved by using Laplace transform techniques. Once we have at hand the solution of the master equation, we can analyze its physicality, i.e. whether the associated map is completely positive and ${\bm \rho}(t)$ is a density operator at all times.
Moreover, it is interesting to compare the solution of this master equation ${\bm \rho}(t)$ for initially factorized states with the tensor product of the two independent system-ancilla evolutions ${\bm \rho}_S(t)\otimes{\bm \rho}_A(t)$, obtained solving the postM master equation for the system in Eq.~(\ref{SLeq}) and von Neumann equation for the ancilla $\dot{\bm \rho}_A(t){=-i}\,\left[\hat H_A,{\bm \rho}_A(t)\right]$, respectively. To this aim we make use of the trace distance $\mathcal{T}({\bm {\bm \rho}},{\bm \varphi}){=}\frac{1}{2}\text{Tr}[|{\bm {\bm \rho}}-{\bm \varphi}|]$ (here $|{\bm A}|{=}\sqrt{{\bm A}^{\dagger} {\bm A}}$ is the trace norm of an arbitrary matrix ${\bm A}$), which defines a distance between the density matrices ${\bm {\bm \rho}}$ and ${\bm \varphi}$.

We start our analysis by considering the system-ancilla initial state ${\bm \rho}(0)=\ket{0}_S\otimes(\ket{0}+\ket{1})_A/{\sqrt{2}}$. Figure~\ref{figura} ({\bf a})  shows the trace distance (dashed line) between ${\bm \rho}(t)$, the solution of Eq.~(\ref{model}), and ${\bm \rho}_S(t)\otimes{\bm \rho}_A(t)$ and the time evolution of the smallest eigenvalue of ${\bm \rho}(t)$ (solid line), for the set of parameters $\nbar{=}1$, $\chi{=}\gamma{=}J/2$. As we can see, not only ${\bm \rho}(t)$ differs substantially from the meaningful expected solution but also violates positivity, being its smallest eigenvalue always negative. This rather surprising result implies that the matrix ${\bm \rho}(t)$ ceases to be a density matrix and that Eq.~(\ref{model}) fails to describe a physical dynamics. As the complete positivity of the map of the single damped qubit (derived by Eq.~(\ref{SLeq})) is guaranteed for any values of the parameters $\nbar$, $\gamma$ and $\chi$, we ascribe such a dramatic implication to the introduction of the ancillary term and check whether there is a regime of the $J$ parameter in which physicality is saved. To this purpose we investigate the evolution of the same system-ancilla factorized initial state as a function of the $J$ parameter. Figure~\ref{figura} ({\bf b}), showing the lowest eigenvalue of the matrix ${\bm \rho}(t)$, demonstrates that positivity is broken even for small value of $J$. This result leaves no doubt about the lack of physicality of the evolution described by the Eq.~(\ref{model}). For the sake of completeness we include the check of the (more restrictive) complete positivity condition. In order to test if a map is CP one can use Choi's theorem~\cite{choi} which states that a map $\Phi$ is CP if the matrix $C_{\Phi}=\sum_{ij}E_{ij}\otimes\Phi(E_{ij})$ is positive, where $E_{ij}$ is a matrix with 1 in the $ij^{th}$ entry and zeros everywhere else. The map $\Phi$ can be determined from the analytical solution (see Appendix). Figure~\ref{figura} ({\bf c}) presents the lowest eigenvalue arising from $C_{\Phi}$ as a function of time and $J$, and confirms that the map loses complete positivity whenever $J\neq0$, as expected.

We notice that the evolution map associated to Eq.~(\ref{model}) is not given by the tensor product of two individual evolution maps of the system and the ancilla, as one would have hoped for. Indeed the overall evolution map is not obtained as the composition of the completely positive maps individually generated by the two terms at  the right hand side of Eq.~(\ref{model}), despite the fact that they do commute and have a common set of
time-independent eigenoperators. In the attempt to trace back to the reasons of the break of positivity and gain insight on such a fictitious evolution, we can calculate the local in time generator of the dynamics. This corresponds to consider the operator ${\bm \rho}(t)$ as the solution of the equation $\dot{{\bm \rho}}(t)=K_{TCL}(t){\bm \rho}(t)$, therefore since we have that ${\bm \rho}(t)=\Phi(t){\bm \rho}(0)$, the local in time generator will be equal to $K_{TCL}(t)=\dot{\Phi}(t)\Phi(t)^{-1}$. 
The expression of $K_{TCL}(t)$, presented in the Appendix demonstrates how the addition of the commutator with the ancillary Hamiltonian leads to the appearance of terms (of the form $\hat{{\bm \sigma}}_i^A\otimes\hat{{\bm \sigma}}_j^B$) correlating the two independent qubits. In this light the trace distance plotted in Fig.~\ref{figura} ({\bf a}) not only describes {\it how far} the evolution map $\Phi(t)$ is from its factorized counterpart, but also accounts for such unphysical correlations created between the qubits.

\section{Introducing an ancillary system: Shabani-Lidar approach} 
As the system dissipator and the ancillary unitary term act on two different Hilbert spaces, the addition of the two independent generators appears as the most natural way to build the master equation of the extended system. Nevertheless, one may ask whether the results obtained are consequences of a possibly naive approach, and an alternative description can cure the physicality of the evolution.
Here we include the ancillary system and its Hamiltonian evolution among the building blocks of the construction of the postM master equation.
 
We take Eq.~(\ref{ShabLidOrig}) and consider $\Lambda$ as the composition of the Markovian dissipative generator $\Lambda_S(t)=e^{\hat{\cal L}_S t}$ acting on the system and the unitary map $\Lambda_A(t)({\bm \rho})=U_A(t)({\bm \rho})U_A^{\dagger}(t)$, with $U_A(t)=e^{-i\, H_A t}$ generated by the Hamiltonian $H_A$. With this substitution and performing the necessary calculations, we obtain the following master equation 
\begin{equation}\begin{split}
\label{pulitaL}
\frac{d}{dt}{\bm \rho}(t)=\hat{\cal L}_S\otimes\hat\openone_A\int^t_0 dt' k(t')e^{\hat{\cal L}_S t'} e^{-i\, H_A t'}{\bm \rho}(t-t')e^{i\, H_A t'}\\ -i\,\left[\hat\openone_S\otimes H_A,\int^t_0 dt' k(t')e^{\hat{\cal L}_S t'} e^{-i\, H_A t'}{\bm \rho}(t-t')e^{i\, H_A t'}\right].
\end{split}\end{equation} 
We start checking the positivity of the evolution of the initial state ${\bm \rho}(0)=\ket{0}_S\otimes(\ket{0}+\ket{1})_A/{\sqrt{2}}$ as a function of the $J$ parameter. The plot in Fig.~\ref{figura} ({\bf d}) shows that positivity is broken also in this case. However, there is an even more important preliminary physicality check that fails to be satisfied: the trace over the system degrees of freedom of Eq.~(\ref{pulitaL}) does not produce the von Neumann equation for the ancilla. To illustrate this statement we consider the simple case where $\gamma=0$ in the dissipator $\hat{\cal L}_S$. For this choice of parameter the first term at right hand side of the master equation in Eq.~(\ref{pulitaL}) is null and, after tracing over the system degree of freedom, one obtains $\dot{\bm \rho}_A{=}-i\,\left[\hat H_A,\int_0^t dt' k(t') e^{-i\, \hat H_A t'} {\bm \rho}_A(t-t') e^{i\, \hat H_A t'}\right]$. This equation cannot be cast in the form of a von Neumann equation, therefore it cannot describe the unitary evolution of the ancillary qubit. Furthermore, such equation does not even describe a physical evolution, since it breaks positivity. The reason for that can be tracked down in the founding idea of the postM approach: the use of an additional measurement on the bath at a random time perturbing the dynamics described by the exact Kraus operator representation. Even if the ancillary system does not interact with the bath, the randomicity of such an instant of time (weighted for the memory function of the bath) modifies the purely unitary dynamics of the ancilla, no longer recovered when tracing over the system. This analysis demonstrates that the {\it verbatim} application of the postM recipe to systems where a partially unitary dynamics is involved risks to produce highly unphysical results. 

\section{Introducing an ancillary system: curing physicality} 
We finally show why the naive generalization of the postM master equation failed. Let us consider Eq.~\eqref{ShabLidOrig} where now $\rho(t)$ is the density matrix of the system-ancilla pair in the interaction picture with respect the bath ($\hat{H}_B$) and system-ancilla ($\hat{H}_0$) free Hamiltonians. We take $\Lambda(t)=\Lambda_S(t)\otimes\openone_A$ with, as before, $\Lambda_S(t)=e^{\hat {\cal L}_St}$. 
%
By performing a transformation to the Schr\"odinger picture by means of the operator $\hat{\cal U}_0(t)=e^{-i\hat{H}_0t}$ and assuming that system and ancilla are not interacting and their initial state is factorized, we get
\begin{equation}
\label{finale}
\begin{aligned}
&\dot{\bm \rho}(t)=-i[\hat H_S,{\bm\rho}_S(t)]\otimes{\bm\rho}_A(t)-i{\bm\rho}_S(t)\otimes[\hat H_A,{\bm\rho}_A(t)]\\
&+\hat{\cal L}_S\int^t_0dt' k(t')e^{\hat{\cal L}_St'}\hat{\cal U}_{S}(t'){\bm\rho_S}(t-t')\hat{\cal U}_{S}^\dag(t')\otimes\bm\rho_A(t),
\end{aligned}
\end{equation}

\noindent where, with a little abuse of notation, the density matrix $\bm \rho(t)$ is meant to be in the Shr\"odinger picture, $\hat{H}_{S,(A)}$ is the free Hamiltonian of the system (ancilla) and $\hat{\cal U}_{S}(t)=e^{-i\hat{H}_St}$.
The presence of the term $e^{\hat{\cal L}_St'}\hat{\cal U}_{S}(t'){\bm\rho_S}(t-t')\hat{\cal U}_{S}^\dag(t')$ tells us that after the extra generalized measurement preformed by the bath at time $t'$ the evolution of the system proceeds for a time interval $t-t'$ under the Markovian map~\cite{ShabaniPRA05}.
The above equation thus generates a global map $\Phi(t)=\Phi_S(t)\otimes\Phi_A(t)$ where $\Phi_S(t)$ is the one induced by the master equation~\eqref{SLeq} (in the Schr\"odinger picture) and $\Phi_A(t)\rho_A\rightarrow\hat{\cal U}_0(t)\rho_A\hat{\cal U}_0^\dag(t)$. The positivity of $\Phi(t)$ is guaranteed by the positivity of the single maps once conditions for complete positivity of $\Phi_S(t)$ are met~\cite{ShabaniPRA05}. $\Phi(t)$ is thus a proper generalization of the postM master equation as one would expect.

In order to derive Eq.~\eqref{finale} {\it ab initio}, one has to propagate the effects of the bath from time $t'$ up to $t$~\cite{ShabaniPRA05}. This obviously makes sense only for those degrees of freedom which are actually interacting with the bath. This is not the case for the ancilla, whose dynamics is entirely unitary. It is then not sufficient that the super operator $\Lambda(t)$ does {\it only} act over the relevant part of the Hilbert space ($\Lambda(t)=\Lambda_S(t)\otimes\openone_A$): the propagation has to involve only that particular subspace. This condition cannot be met if one derives Eq.~\eqref{SLeq} in the Schr\"odinger picture by simply replacing $\Lambda(t)$ , given that the memory kernel ``keeps track'' of all of the degrees of freedom regardless wether or not they did interact with the bath at one point in the past. This problem can be overcome by introducing a memory kernel which is of the form
$k(t)\openone_M\otimes\delta(t)\openone_F$ where $\openone_M$ is the identity operator over the subspace the bath ``has memory of'' whereas $\openone_F$ does act over the ``free'' subspace.

\section{Conclusion and final remarks} 
In this paper we highlighted the hazard of using postM master equations in systems containing a partially unitary dynamics. The specific example of a physical postM master equation describing a damped qubit extended to include an ancillary one shows that the introduction of a Hamiltonian term has drastic effects even when the ancilla is unitarily evolving and totally decoupled from the system and the environment. Both the intuitive approach, based on the summation of the two independent generators of system and ancilla, and the postM derivation, accounting for the ancilla degree of freedom {\it ab initio}, break fundamental physicality requirements such as positivity. As final remarks we add that the same conceptual problems arise when using Hamiltonian terms coupling system and ancilla (such as Ising coupling), and also in the case where no ancilla is brought in and the Hamiltonian term acts on the system only. Analogous results can be found when the equation to generalize is a phenomenological memory-kernel master equation. This difficulty has also been considered in a different context in~\cite{budini}, where complete positivity for an initially well defined integrodifferential master equation has been preserved upon the introduction of a Hamiltonian term only by suitably modifying the integral kernel. Due to their very construction, those master equations do not always guarantee positivity and CP, even in the single-qubit formulation. 

We have addressed the use of postM master equations in the case of composite (interacting and non-interacting) systems, pointing out that quite a careful approach should be taken to include the unitary part of the evolution in the measurement-based picture at the basis of the postM framework. For the case of non-interacting particles, only one of which experiencing the effects of the postM environment, the intuitive combination of the separate dynamics of the two systems may lead up to unphysical results. Our formal finding, which are interesting from a fundamental viewpoint, will have implications at the pragmatic side in schemes where the ancilla is used in order to probe the properties of the the environment into which the system is immersed~\cite{CampbellPRA10}.


\acknowledgments
This work was supported by the NI DEL, the Magnus Ehrnrooth Foundation, MIUR under PRIN 2008 and COST under MP1006, SFI under grant numbers 05/IN/I852 and 05/IN/I852 NS, IRCSET though the Embark initiative (RS/2000/137) and the UK EPSRC (EP/G004759/1). 

\renewcommand{\theequation}{A-\arabic{equation}}
\setcounter{equation}{0}  
\section*{APPENDIX A - Derivation of the time-local generator}  

Here we present the derivation of the time-local generator associated to the master equation in Eq. (4). By using the decomposition of  ${\bm \rho}(t)$ in terms of the damping basis $\{\hat{{\bm \sigma}}_i^S\}~(i=0,3,\pm$ with $\hat{\bm \sigma}^{S}_3$ the $z$ Pauli matrix) given by ${\bm {\bm \rho}}(t)=\sum_{i,j}c_{ij}(t) \hat{{\bm \sigma}}_i^S\otimes\hat{{\bm \sigma}}_j^A$, we can derive the following set of differential equations
\begin{equation}
  \dot{ c}_{i j} (t)=- i \sum_{k l} h_{i j}^{k l} c_{k l} (t) +
  \lambda_i \int_0^t d \tau k (\tau) e^{\lambda_i \tau} c_{i j} (t - \tau) 
  \label{eq:sist}
\end{equation}
with $h^{k l}_{i j}\equiv \text{Tr} \left\{ {\hat{\check{\bm \sigma}}_i^S \otimes \hat{\check{\bm \sigma}}}_j^A \left[ H, \hat{{\bm \sigma}}_k^S\otimes\hat{{\bm \sigma}}_l^A \right] \right\}$ and $\hat{\check{\bm \sigma}}_i^{S,A}$ the elements of the dual damping basis for the system and ancilla. These equations can be solved in the Lapace space as
\begin{equation}
  c_{i j} (t)=\text{Lap}^{- 1} \left[ \frac{1}{s + i h_j -
  \frac{\lambda_i \chi}{\chi + s - \lambda_i}} \right] c_{i j} (0) . 
  \label{eq:aijt}
\end{equation}
where, according to $\hat H$ as given in Eq.~(5), one has $h^{k l}_{i
j} =: \delta_{i k} \delta_{j l} h_j$ and the time-dependence of the Laplace anti-transform $\text{Lap}^{-1}[\cdot]$ is understood.
The solution shows that the evolution map is diagonal in the basis $\left\{\hat{{\bm \sigma}}_i^S\otimes\hat{{\bm \sigma}}_j^A\right\}$ 
and can be written as a diagonal $16\times16$ matrix 
\begin{eqnarray}
\Phi(t)= \text{diag} \left\{ \zeta_{i j} (t) \right\}  \label{eq:f}
\end{eqnarray}
with elements
\begin{equation}\begin{split}
  \zeta_{i j} (t) = e^{- \Omega_{i j} t / 2} \left[\cosh \left( \frac{\Delta_{i j} t}{2} \right) +
  \frac{\Omega^{\ast}_{i j}}{\Delta_{i j}} \sinh \left( \frac{\Delta_{i j} t}{2}
  \right) \right],\label{eq:zetaij}
\end{split}\end{equation}
where $ \Delta_{i j}{=} \sqrt{(\chi + \lambda_i)^2 - h^2_j - 2 i h_j (\chi - \lambda_i)}$ and $\Omega_{i j}{=}\chi{-}\lambda_i{+}i h_j$. Evidently, the coefficients $\zeta_{i j} (t)$ cannot be written as $\zeta_i (t) \zeta_j(t)$, therefore the evolution map $\Phi(t)$ is not given by the tensor product $\otimes_{k=S,A}\Phi_k (t)$ of single-qubit maps $\Phi_k (t)$.
Having the matrix representation of the evolution map, we can calculate the time-local generator $K_{\text{TCL}} (t) = \dot{\Phi} (t)
\Phi(t)^{- 1}$, whose matrix representation in the two-qubit damping basis 
is given by
\begin{equation}
  K (t) = \dot{\Phi}(t)\Phi(t)^{- 1} = \text{diag} \left\{
  \frac{\dot{\zeta_{i j}} (t)}{\zeta_{i j} (t)} \right\} .  \label{eq:ktcl}
\end{equation}
Given a basis $\left\{ \hat B_a \right\}_{a = 0, \ldots,
N^2-1}$ in $\mathfrak{L}( \mathcal{H}^N)$, one can define the scalar product between two super-operators 
defined in the space of linear operators $\mathfrak{L}( \mathcal{H}^N)$~\cite{GoriniJMP76} as
\begin{equation}
\begin{aligned}
  \langle \hat\Lambda_1, \hat\Lambda_2 \rangle & =  \sum_a \text{Tr} \left\{
  \hat\Lambda_1 [\hat B_a]^{\dag} \hat\Lambda_2 [\hat B_a] \right\} .  
  \end{aligned}
  \label{eq:scp}
\end{equation}
A basis in the space of super-operators is given by the family $\left\{ \hat\varphi_{\beta \gamma} \right\}_{\beta, \gamma = 0,
\ldots, N^2-1}$ such that
$ \hat\varphi_{\beta \gamma} [\rho] = \hat B_{\beta} \varrho
 \hat B^{\dag}_{\gamma}$ with $\varrho$ a generic density matrix. Therefore, any super-operator $\hat\Lambda$ acting on $\mathfrak{L}( \mathcal{H}^N)$ can
be written as
\begin{equation}
 \hat \Lambda \varrho= \sum_{\beta \gamma} \Lambda_{\beta \gamma} \hat\varphi_{\beta
  \gamma} [\rho] = \sum_{\beta \gamma} \Lambda_{\beta \gamma} \hat B_{\beta}
  \varrho  \hat B^{\dag}_{\gamma},  \label{eq:lambdav}
\end{equation}
with
$\Lambda_{\beta \gamma}{=}\langle \varphi_{\beta \gamma}, \Lambda_{}
  \rangle$. 
We take a basis $\left\{\hat B_a \right\}_{a = 0, \ldots N^2 - 1}$ such that $\hat{B}_0 = \frac{1}{\sqrt{N}}\hat\openone$ and $\text{Tr} \{
\hat{B}_a\} = 0$, $\forall{a} = 1, \ldots, N^2 - 1$. From the
expression
\begin{equation}
  K_{\text{TCL}} (t) [\rho]=\sum^{N^2 - 1}_{\beta, \gamma = 0} K_{\beta
  \gamma} (t) \hat{B}_{\beta} \rho \hat{B}^{\dag}_{\gamma} 
  \label{eq:krho}
\end{equation}
of the time-local generator, we obtain its {\it Lindblad-like version} simply by
removing the first row and the first column of the corresponding matrix of coefficients, since the evolution map is trace-preserving and thus ${\rm Tr}\{K_{TCL}(t)[\rho]\} = 0$. We thus obtain
\begin{equation}\label{eq:lindb}
  K_{\text{TCL}} (t) [\rho]=- i \left[ \hat H', \rho \right] + \sum^{N^2 -
  1}_{\beta, \gamma = 1} K_{\beta \gamma} (t) \left( \hat{B}_{\beta}
  \rho \hat{B}^{\dag}_{\gamma} - \frac{1}{2} \left\{
  \hat{B}^{\dag}_{\gamma} \hat{B}_{\beta}, \rho \right\}
  \right), 
\end{equation}
with the Hamiltonian term given by $\hat H'=\frac{1}{2 i} (\hat g^{\dag} - \hat g)$ and
\begin{equation}\label{eq:hgorini}
\hat g=\frac{1}{\sqrt{N}} \sum^{N^2 - 1}_{\beta = 1} K_{\beta 0} (t)\hat B_{\beta}. 
\end{equation}
In the basis given by
\begin{equation}\begin{split}
\hat {B}^{SA}_0&=\hat\openone^{SA}/ 2\\
\hat {B}^{SA}_a&=(\hat\openone^{S} \otimes \hat{\bm \sigma}^{A}_j) / 2 \hspace{2em} j = 1,2, 3 ;
  \, a = 1,2, 3\\
\hat {B}^{SA}_a&=(\hat{\bm \sigma}^{S}_i \otimes \hat\openone^{A}) / 2 \hspace{2em} i = 1,2, 3 ;
  \, a = 4, 5, 6\\
\hat {B}^{SA}_a &=(\hat{\bm \sigma}^{S}_i \otimes \hat{\bm \sigma}^{A}_j) / 2 \hspace{2em} i,j = 1,2, 3 ; 
\, a = 7, \ldots, 15,  \label{eq:basis}
\end{split}
\end{equation}
where $\hat{\bm \sigma}^{S,A}_{1, 2}$ are the $x$ and $y$ Pauli matrices respectively, the local generator in Lindblad form is given by Eq. (\ref{eq:lindb}) with
\begin{equation}
  K_{\beta \gamma} (t)=\sum_a \text{Tr} \left\{\hat{B}_{\gamma}
 \hat{B}^{\dag}_a\hat{B}^{\dag}_{\beta} K_{\text{TCL}} [
  \hat{B}_a] \right\}  \label{eq:kbj}
\end{equation}
with $K_{\text{TCL}} (t) [ \hat{B}_a]$ that can be explicitly calculated using the relation
\begin{equation}
  K_{\text{TCL}} [\rho] =  \sum_{i j} \frac{\dot{\zeta_{i j}}
  (t)}{\zeta_{i j} (t)} \text{Tr} \left\{ \hat{\check{\bm \sigma}}_i^S \otimes \hat{\check{\bm \sigma}}_j^A \rho \right\} \hat{{\bm \sigma}}_i^S
  \otimes \hat{{\bm \sigma}}_j^A~~(i,j=0,z,\pm) .  \label{eq:ktclrho2}
\end{equation}
The Hamiltonian term is obtained substituting the expression $  K_{\beta 0} (t)= \frac{1}{2} \sum_a \text{Tr} \left\{
  \hat{B}^{\dag}_a \hat{B}^{\dag}_{\beta} K_{\text{TCL}} [
  \hat{B}_a] \right\}$ in Eq.~(\ref{eq:hgorini}).
Due to the particular choice of the basis in the space of the super-operators, the dynamics can create correlations between the
two qubits only if there are non-zero coefficients $K_{\beta \gamma}$ for
$\beta, \gamma = 7, \ldots, 15$, {\it i.e.} if the dynamics involves basis elements of the form
$\hat{\bm \sigma}^S_i \otimes \hat{\bm \sigma}^A_j$. Thus, a necessary condition for the generation of
correlations is that there are non zero values outside the $7 \times 7$
top-left corner reduction of the coefficients matrix given by $K_{\beta
\gamma}$. For example, one can directly check that if $\zeta_{i j} (t) =
\zeta_i (t) \zeta_j (t)$ then $K_{\beta \gamma} = 0$ for $\beta, \gamma = 7,
\ldots 15$, while this is not the case for $\zeta_{i j} (t)$ given by
Eq. (\ref{eq:zetaij}). In this case, there are non-zero coefficients related to
$\hat{\bm \sigma}^S_i \otimes \hat{\bm \sigma}^A_j$ in the dissipative part of the generator and there
is also a non zero coefficient associated to $\hat{\bm\sigma}^S_z \otimes \hat{\bm\sigma}^A_z$ in
the Hamiltonian part. This means that the Hamiltonian term added in the
master equation given by Eq. (4) modifies entirely
the structure of the time-local generator.

\renewcommand{\theequation}{B-\arabic{equation}}
\setcounter{equation}{0}  
\section*{APPENDIX B - Ab initio extension of the post-Markovian master equation}  

Here we present the mathematical calculations we carry over to derive the Master Equation in Eq. (7), obtained by including ab initio the ancillary system and its Hamiltonian evolution. We recall the notation used: the Markovian dissipative generator is $\Lambda_S(t)=e^{\mathcal{L}_S t}$, while the unitary map acting on the ancilla is  $\Lambda_A(t)[{\bm \rho}]=U_A(t)[{\bm \rho}]U_A^{\dagger}(t)$. The starting point is Eq. (1), namely
\begin{equation}
\frac{\partial\rho}{\partial t}=\int_0^t dt' k(t')\Lambda(t') \frac{\partial \Lambda(t-t')}{\partial (t-t')}\Lambda^{-1}(t-t') \rho(t-t').
\end{equation}
As the system and ancilla are not interacting we set $\Lambda(t)=\Lambda_S(t)\otimes\Lambda_A(t)$. We can thus write
\begin{equation}
\Lambda^{-1}(t-t'){\bm \rho}(t-t')=\Lambda_S^{-1}(t-t') U_A^{\dagger}(t-t'){\bm \rho}(t-t')U_A(t-t'),
\end{equation}
and thus, given that $\frac{\partial \Lambda_s (t-t')}{\partial (t-t')} = \mathcal{L}_S e^{\mathcal{L}_S(t-t')}=\mathcal{L}_S\Lambda_S(t-t')$, we have that
\begin{equation}
\dot{\Lambda}(t-t')\Lambda^{-1}(t-t'){\bm \rho}(t-t')=\mathcal{L}_S\rho(t-t')-i [H_A,{\bm \rho}(t-t')].
\end{equation}
Using the previous results we finally obtain
\begin{equation}\begin{split}
\frac{d}{dt}{\bm \rho}(t)=\hat{\cal L}_S\otimes\hat\openone_A\int^t_0 dt' k(t')e^{\hat{\cal L}_S t'} e^{-i\, H_A t'}{\bm \rho}(t-t')e^{i\, H_A t'}\\ -i\,\left[\hat\openone_S\otimes H_A,\int^t_0 dt' k(t')e^{\hat{\cal L}_S t'} e^{-i\, H_A t'}{\bm \rho}(t-t')e^{i\, H_A t'}\right].
\end{split}\end{equation}

\end{document}